\documentclass[conference]{IEEEtran}

\usepackage{cite}
\usepackage{amsmath,amssymb,amsfonts}

\usepackage{algorithmic}
\usepackage{graphicx}
\usepackage{textcomp}
\usepackage{euscript}
\usepackage{xcolor}
\usepackage{booktabs}
\usepackage{float}
\usepackage[T1]{fontenc}
\usepackage{epstopdf}
\pdfoutput=1
\begin{document}
\IEEEoverridecommandlockouts
\title{Power Allocation for Device-to-Device Interference Channel Using Truncated Graph Transformers
\thanks{This work was supported by The Hong Kong University of Science and Technology (HKUST) Startup Fund under Grant R9249.}}

\author{\IEEEauthorblockN{Dohoon Kim and Shenghui Song}
\IEEEauthorblockA{Dept. of ECE, The Hong Kong University of Science and Technology, Hong Kong\\
Email: \{dhkimac, eeshsong\}@ust.hk}}

\maketitle

\begin{abstract}
Power control for the device-to-device interference channel with single-antenna transceivers has been widely analyzed with both model-based methods and learning-based approaches. Although the learning-based approaches, i.e., data-driven and model-driven, offer performance improvement, the widely adopted graph neural network suffers from learning the heterophilous power distribution of the interference channel. In this paper, we propose a deep learning architecture in the family of graph transformers to circumvent the issue. Experiment results show that the proposed methods achieve the state-of-the-art performance across a wide range of untrained network configurations. Furthermore, we show there is a trade-off between model complexity and generality.
\end{abstract}


\section{Introduction}
Power allocation is a fundamental problem for wireless communications. For device-to-device (D2D) systems where single antenna transmitter-receiver pairs communicate in an interference channel, the problem can be abstracted into a sum-rate maximization with power constraints. However, the non-convex and NP-hard nature makes optimizing this simple formulation a challenge \cite{luo_dynamic_2008}. There have been various approaches to approximate a solution, such as solving it on the Lagrangian dual domain \cite{yu_dual_2006} or weighted minimum mean squared error (WMMSE) minimization \cite{shi_iteratively_2011}. These classical methods guarantee convergence to a local minima, but are computationally complex and not suitable for large scale wireless networks. Furthermore, the iterative nature of these algorithms prevent the utilization of parallel computing.\par
Recent advances in deep learning has inspired data-driven approaches to tackle this optimization problem \cite{chowdhury.etal_2022,9944643,eisen_large_2019}. These methods outperform classical algorithms with a fraction of computational complexity, in exchange for generality. Purely data-driven models can not guarantee performance outside of its training data distribution. Model-driven approaches merge classical algorithms with deep learning models through algorithmic unfolding to utilize the strong prior of conventional techniques \cite{chowdhury.etal_2022}. The deep learning model is trained to accelerate the convergence of existing algorithms, drastically reducing the iteration count. However, such an unfolding algorithm retains the computational complexity of the classical algorithm it is optimizing, and retains the iterative nature. \par
For both data-driven and model-driven methods, graph neural networks (GNNs) have been the architecture of choice due to their strong prior on graph data structures. GNN's kernel sizes are constant to arbitrary size of graphs, and its permutation invariance reduces data complexity during learning \cite{9944643}.  Since wireless networks can be easily modeled as a graph, where individual devices can be treated as nodes and communication links as edges, GNNs can be easily adapted for wirless network datasets. 
However, we would like to argue that the basic GNN architecture is not suitable for the concerned power allocation problem, as the interference channel is always a fully connected graph, and its optimal power distribution is heterephilous, which GNN models suffer from learning \cite{liu2019hyperbolic}.\par

In this work, we explore a different model architecture, i.e., transformers. Transformers have achieved great successes in the area of natural language processing (NLP), where large pretrained models could be fine tuned on various tasks. Models like GPT-3 or Lambda are giant transformer networks with billions of parameters that generate state-of-the-art performance across different NLP domains \cite{lin2022survey}. Their attention mechanism maps a weight to all input nodes in the sequence, which allows gradient flow on long dependencies of a sentence. As attention maps a scalar weight to all words in a sentence, it is equivalent to providing an edge weight to a fully connected graph with the same number of nodes as the input sequence.\par

In this paper, we propose to revise the graph transformer architecture for the power allocation problem in the D2D interference channel. Graph transformers extends the concept of attention weights from transformers to GNNs, where it generates an attention map that acts as a soft weight across all neighbors \cite{dwivedi_generalization_2021}. Extensive experiments validate that the proposed architecture outperforms classical algorithms and other learning based models across different network topologies and configurations with just a single trained model. We further show that model complexity correlates to model generalization, which proposes an interesting trade-off between the computational complexity during inference and the frequency for model retraining.

\section{Problem Formulation and Preliminaries}
\subsection{Weighted Sum-Rate Maximization}
Consider a single-input single-output (SISO) ad hoc interference network with $n$ transmitter-receiver pairs. Denote $h_{ij}$ to be the channel coefficient between transmitter $i$ and receiver $j$, $p_i$ as the power allocated to transmitter $i$, and $\sigma^2$ as the additive white Gaussian noise (AWGN) power level. The channel rate of the $i$-th transmitter-receiver pair can be expressed as:

\begin{equation}
    c_i = log_2\left(1 + \frac{|h_{ii}|^2 p_i}{\sigma^2 + \sum_{j=1}^{j\neq i} |h_{ij}|^2 p_j}\right), i=1,...n.
\end{equation}

The design objective is to maximize the sum rate across all pairs under the maximum power constraint. Additionally, each pair would have a weight $w_i$ on the channel rate to enforce priority or fairness across devices. The constrained weighted sum rate maximization problem can be written as:
\begin{align}
    \max_{\textbf{p}}&  \sum w_{i}c_i, \\
    \text{s.t }& \textbf{p} \in [0,P_{max}]^n
\end{align}
Our goal is to find a data-driven model that determines the maximizing power $\textbf{p} \in \mathbb{R}_+^n$ given an instantiated channel state $\textbf{H}  \in \mathbb{R}_+^{n \times n}$with  $\textbf{H}_{i,j} = h_{ij} $ and noise power $\sigma^2$. \par
\subsection{Graph Neural Networks}
GNN is a family of neural network architectures designed to be trained on graph data structures. A single pass of a GNN layer consists of three stages: message formulation, neighbor aggregation, and feature update. The message formulation function $m(\cdot)$ creates a message feature vector, generally using the corresponding node embedding $\textbf{x}_i$ and edge embedding $\textbf{e}_{ij}$. Neighbor aggregation $\bigoplus$ is a symmetric operation across all neighbors of a specific node, which grants permutation invariance to the input sequence. $j \in \EuScript{N}(i)$ is a set of nodes that are connected with node $i$ in the graph. Generally, these would be summation, average, or maximum, but any other symmetrical operations would suffice. The authors of \cite{corso2020principal} discussed the efficacy for various candidates, and showed that no specific operation is superior and subject to case by case identification. Finally, the feature update $\sigma(\cdot)$ takes both the aggregated message and its node feature embedding to compute the new embedding for node i. 
\begin{equation}
    \textbf{z}_i = \sigma(\textbf{x}_i, \underset{j \in \EuScript{N}(i)}{\bigoplus}m(\textbf{x}_i, \textbf{x}_j, \textbf{e}_{ij})).
\end{equation}
Note that if $\bigoplus$ is a sum operation, this is equivalent to multiplying the node feature matrix $\textbf{X} \in \mathbb{R}^{n \times d}$ with the adjacency matrix $\textbf{A} \in \mathbb{N}_0^{n \times n}$, where the value of $\textbf{A}_{ij} = 1$ if node i and j are connected, and 0 otherwise. In this case, we can imagine a basic message passing GNN as a masked linear layer in relation to the adjacency matrix.
Various architectures of GNN have been utilized in the literature for the power allocation problem. The authors of \cite{chowdhury.etal_2022} proposed a GNN to accelerate WMMSE, and \cite{9944643} utilized a message passing GNN where the message is generated with concatenated node and edge feature embeddinigs .\par
While GNNs show better expressitivity in graph data, they have their limits. The limit relevant to our problem formulation is their strong bias on homophily, which means nodes connected with each other will have the same label. Detailed analysis in regards to the power control problem will be discussed in Section V-C.

\subsection{Transformers}
Transformer is another family of neural network architectures that was developed for NLP problems \cite{vaswani2017attention}. Its attention mechanism was devised to circumvent the problem of gradient vanishing in long sentences. Information of words at the start of the sentence would not pass through the model when inferring on the words on the later part. Attention solves this by placing a soft weight on all words in the sentence for each step, which is calculated on run-time. The attention map is calculated using a query, key, and value matrix. If the model has an input feature vector $\textbf{x}_\textbf{i} \in \mathbb{R}^{n \times f}$, the attention module would contain linear weight $\textbf{W}_\textbf{q} \in \mathbb{R}^{d \times n}, \textbf{W}_\textbf{k}  \in \mathbb{R}^{d \times n}, \textbf{W}_\textbf{v}  \in \mathbb{R}^{d \times n}$ which would compute the feature matrix $\textbf{Q}, \textbf{K}, \textbf{V} \in \mathbb{R}^{d \times f}$. When multiple attention modules are stacked, this is called a multi-head attention. The attention module could be expressed as:
\begin{equation}
    \textbf{z}_i = \text{softmax}(\frac{\textbf{Q}\textbf{K}^T}{\sqrt{d}})\textbf{V}.
\end{equation}
Because attention is individually computed for each word, and the matrix multiplication of the attention weights to the value matrix is equivalent to a sum neighbor aggregation, transformer models are inherently permutation invariant. Note that the input that generates the $\textbf{Q}$ ,$\textbf{K}$, and $\textbf{V}$ can be from different input vectors. This is called cross-attention, which would be used for our proposed method as the query comes from the target node, and the keys and values come from the neighboring nodes. 

\section{Truncated Graph Transformers (TGTs)}
\subsection{Model Architecture}
\begin{figure*}[tb]
\centerline{\includegraphics[width= 0.85\textwidth]{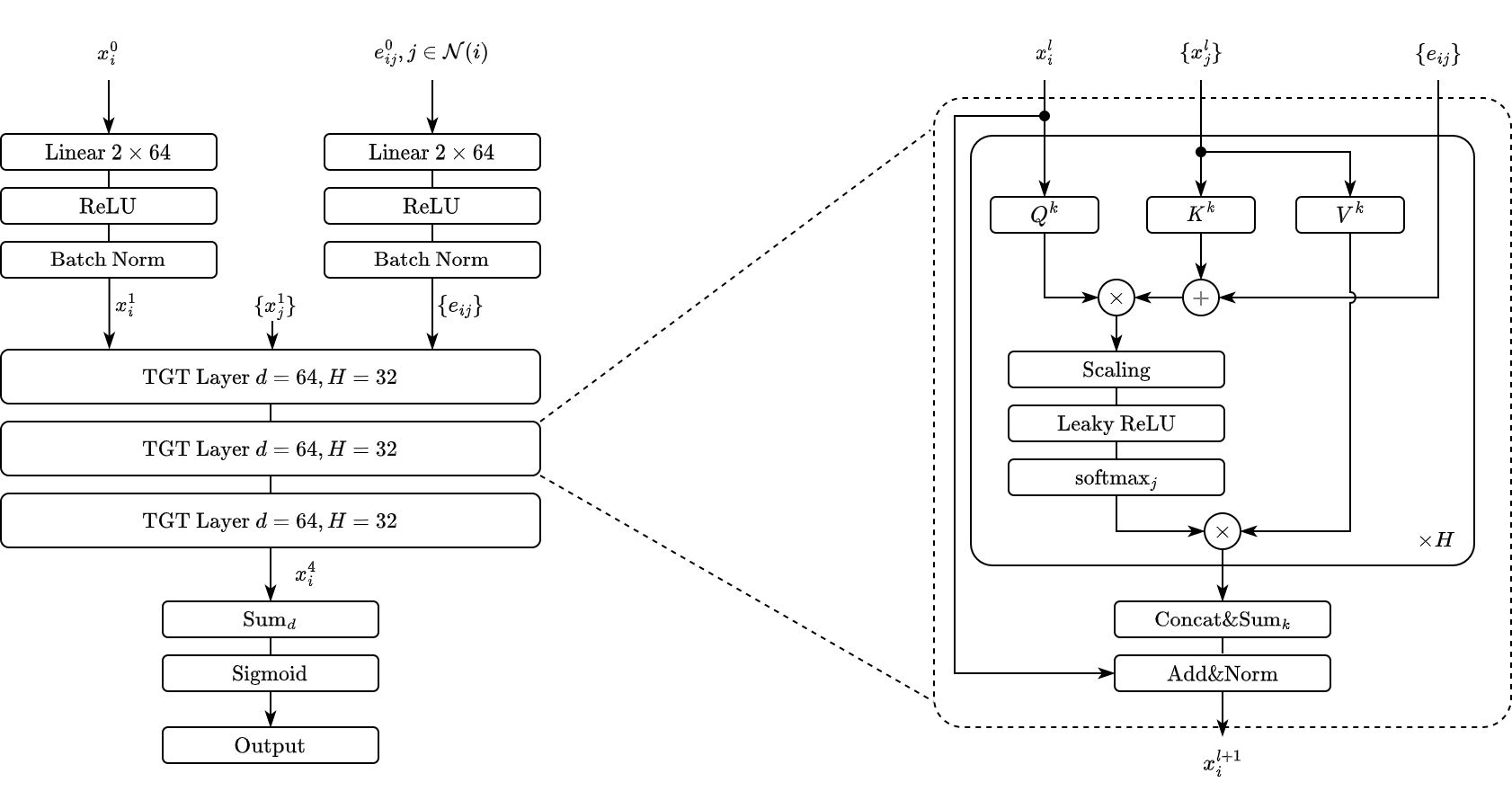}}
\caption{Block diagram of the TGT algorithm. Note that the Transformer layer weights are shared across the three instances.}
\label{fig:dia}
\end{figure*}
Transformers take a sequence input of arbitrary length and compute a permutation invariant operation. This can be alternatively viewed as a GNN on a fully connected simple graph. The authors of \cite{dwivedi_generalization_2021} generalized the transformer to a GNN framework, named graph transformers. Details of the structure is nearly identical with the original self attention, but the difference comes from how the input data is handled. Here, the queries and keys are generated from the corresponding node embeddings, and the attention is mapped across the neighbors. A possible integration of edge feature vector to the process is also introduced. \par
We revise the graph transformer architecture discussed in \cite{dwivedi_generalization_2021} for better convergence on the power allocation problem. Fig. \ref{fig:dia} illustrates the proposed model. First, both the node and edge features are passed through a fully connected layer with batch normalization into a higher dimension embedding with dimension $d$. The embedding is then fed through a multi-head cross attention 3-layer model defined as:
\begin{align} 
    &\hat{\textbf{x}_i}^{l+1} = \sum_{k=1}^{H} \underset{j \in \EuScript{N}(i)}{\sum}w^{kl}_{ij}\textbf{V}^k\textbf{x}^{kl}_j, \\
    \text{where, }& \nonumber \\
    &w^{kl}_{ij} = \text{softmax}_j(\text{LeakyReLU}(\frac{\textbf{Q}^kx^{kl}_i(\textbf{K}^{k}\textbf{x}^{kl}_j + \textbf{e}^k_{ij})}{\sqrt{d}})).
\end{align}

\noindent Here, $\textbf{x}^l_{ij}, \textbf{e}_{ij} \in \mathbb{R}^{d}$ are the node and edge embeddings, $\textbf{Q}^k, \textbf{K}^k, \textbf{V}^k \in \mathbb{R}^{d/H \times d/H}$ are the query, key, and value weights of the $k^{th}$ head, and $H$ is the number of heads for the multi-head attention. $w_{ij}^{kl}$ is the attention map for the $ij$-th message on head $k$, layer $l$. The right side of Fig. \ref{fig:dia} illustrates how a single TGT layer pass is computed. First $\textbf{x}^l_{ij}, \textbf{e}_{ij}$ are split into $\textbf{x}^{kl}_{ij}, \textbf{e}^k_{ij} \in \mathbb{R}^{d/H}$ for each head. Then the query is generated from the target node and the keys are generated from its neighboring nodes. The edge information is injected as a sum to the key feature vector. The product of the query and keys are scaled, and passes through a Leaky ReLU before going through softmax \cite{velivckovic2017graph}. Ablation study showed that models with no shared weights only provide marginal improvement on performance while tripling the number of trainable parameters. Thus, $\textbf{Q}^k, \textbf{K}^k, \textbf{V}^k $ are shared across all three layers. The output of the multi-head attention is followed by a layer normalization \cite{ba2016layer} with a skip connection of the previous layer output, as:
\begin{equation}
    \textbf{x}_i^{l+1} = LayerNorm(\hat{\textbf{x}_i}^{l+1} + \textbf{x}_i^l).
\end{equation}
The final layer of the model is just a sum along the feature dimension to flatten the node feature to a scalar, which is then passed through a sigmoid activation to limit the range of the output to $[0,1]$. This can be expressed as:
\begin{equation}
    p_i = \frac{1}{1+exp(-(1,1,...,1)\textbf{x}_i^L)}.
\end{equation}
As our model reduces the number of non-linearities from the original model, we name it as the Truncated Graph Transformer (TGT).
\subsection{Graph Homophily }
In this section, we attempt to explain why TGT outperforms basic GNNs by showing that the power control problem with low noise is heterophilous. Graph homophily is an index that measures whether a node with an adjacent edge would have the same label. We start with the index definition by \cite{pei2020geom}:
\begin{equation}
h=E[\frac{1}{d_i}\sum_{j \in  \EuScript{N}(i)} |x_i = x_j|^\textbf{1} ].
\end{equation}
Where $||^\textbf{1}$ is a function that returns 1 if the two values are the same, and $d_i$ is the degree of node $i$. $h$ would map to a value $[0,1]$, where 1 represents that the graph is highly homophilic. Citation graphs or relation graphs are known to be homophilic, as papers with similar topics would be cited together and similar people will be in a closer relationship. GNNs are known to better perform in homophilic datasets, which is due to their neighbor aggregation design on a fixed graph structure. The author of \cite{wu2019simplifying} showed that when removing the non-linearity of the GNN, it can be abstracted into a Chebyshev low pass filter on the graph, which indicates that the performance gained from utilizing GNNs may not be from the non-linear feature extractions. In order to circumvent this issue, researchers are exploring more sophisticated GNN architectures that include sub-graph information \cite{frasca_understanding_2022}, sheaf diffusion \cite{bodnar_neural_2022}, and mapping the graph into a hyperbolic space \cite{liu2019hyperbolic}.\par
Due to the network modeled in the problem formulation is a fully connected graph with a scalar edge weights and power allocation, we expand the definition of graph homophily for a graph with scalar edge weight $e_{ij}$ and node label $x_{i} \in [0,1]$ as:
\begin{equation}
h=E_{i}[\frac{1}{\sum_{j \in  \EuScript{N}(i)} e_{ij}}\sum_{j \in  \EuScript{N}(i)} e_{ij}(1 -| x_i - x_j|) ].
\end{equation}

\begin{table}[tb]
\caption{Homophily $h$ on a wireless network of 50 pairs}
\centering
\begin{tabular}{llllllll}
\toprule
  & \multicolumn{7}{c}{Noise Power Level $\sigma^2$}                       \\ 
  & $10^{-6}$     & $10^{-5}$     & $10^{-4}$      & $10^{-3}$      & $10^{-2}$      & $10^{-1}$      & $10^{0}$      \\ \hline
$h$ & 0.417 & 0.418 & 0.468 & 0.560 & 0.688 & 0.817 & 0.914 \\ 
\bottomrule
\end{tabular}
\label{tab:ho}
\end{table}
\noindent Table \ref{tab:ho} shows the homophily of a network channel \textbf{H} with 50 pairs, where the optimal power generated by WMMSE is considered as the label. As the noise level increases, the allocated power changes from a heterophilous to a homophilic distribution. Intuitively, the power control problem for the SISO interference channel can be abstracted into finding the optimal volume for each person in a busy dinner party. If two people were close together with their partners in equal distance, it would be beneficial for one to remain silent and the other to speak with their loudest voice, instead of the two speaking over each other. With increased noise power, the optimal power allocation saturates to full power, as the noise power level overpowers the interference from neighboring pairs.\par
Graph Network Transformers are more resilient to heterophilous datasets due to the attention weights acting as active graph rewiring. The wireless network is formulated as a graph with soft neighbor connections, in which every node is connected with a scalar weight associated to it. By applying attention to each neighbors, the model is effectively rewiring the graph each layer, avoiding the over smoothing issue of GNNs when aggregating with a fixed graph topology.

\section{Numerical Experiments}
\subsection{Performance on Singular Network Size}
We perform experiments on a simulated SISO D2D network over an AWGN channel of fixed noise $\sigma^2 = 2.6 \times 10^{-5}$. $n$ transmitters are uniformly distributed on a square plane with range $[-n,n]^2$, where the receiver is placed within the distance of $[1,n/4]$ around the target transmitter. A lower bound on the receiver location is placed to prevent numerical instability when calculating the path gain. The channel is constructed with a constant path-loss defined as $h^p_{ij} = ||t_i - r_j||^{-2.2}$, where $t_i, r_j$ are the coordinates of the corresponding transmitter and receiver. We then multiply with the large-scale channel coefficient coefficient $h^f_{ij} \sim \text{Rayleigh}(1)$. This specific problem setting has been used across various data driven power allocation publications, and has been chosen here for ease of comparison across different methods \cite{chowdhury.etal_2022,eisen_large_2019}. We assume equal weight $w_{i}=1$ for all pairs.\\
For all data-driven benchmarks, the models were trained on a dataset of 25,000 network instances, which consists of 50 randomly sampled channel states on a 500 randomly sampled network topology. For the data-driven benchmarks, published configurations for model dimension and training strategies were retained, but the learning rate were adjusted so that the model could stably converge due to difference in simulated conditions. Then the model was tested on previously unseen 1000 network instances with 50 randomly sampled channel states on  50 randomly sampled network topologies. \par
The proposed model, TGT, is a 3 layer graph transformer model with 32 head attention, where the embedding dimension $d = 64$. As the channel $h_{ij}$ is a near zero value, we normalize the channel matrix \textbf{H} to a range of $[0,1]$. The model was trained using a AdamW\cite{loshchilov2017decoupled} optimizer with learning rate set to $5 \times 10^{-4}$. All models were trained for 50 epochs.
The channel state $\textbf{H}$ needs to be represented as a graph data to be processed by a GNN. For our method, we convert $\textbf{H}$ to an undirected graph $G$ with $n$ nodes. In particular, $G$ is a fully connected graph with self loops where feature vector of the $i$-th node and the $ij$-edge are given by $\textbf{x}_i^0 = [h_{ii}, w_{i}] \in \mathbb{R}^{2}$ and $\textbf{e}_{ij} = [h_{ij},h_{ji}] \in \mathbb{R}^2$, respectively. Implementation details can be accessed through the github repository \cite{Dohoon2023}.  \par
We perform numerical comparison across the following benchmarks:
\begin{itemize}
    \item Max Power is a naive strategy where all transmitters transmit maximum power $P_{max} = 1$.
    \item WMMSE\cite{shi_iteratively_2011} acts as a baseline for the classical method, which is run for 100 iterations.
    \item PCGNN\cite{9944643} utilizes a variant of a GNN with its edge weights embedded onto the message. It achieves state-of-the-art performance on complex interference channels.
    \item UWMMSE\cite{chowdhury.etal_2022} is a model-driven approach that utilizes GNNs to accelerate the conversion of WMMSE with only 4 iterations. It achieves state-of-the art performance on a fixed network topology.
    \item Graph Trasformer \cite{dwivedi_generalization_2021} is run with the same number of layers, heads, and $d$ to compare with the proposed method. Note that this model has 173K trainable parameters, while ours only have 13.1K.

\end{itemize}

\begin{table}[tb]
\caption{Experiment results on a various range of network size.}
\centering
\begin{tabular}{@{}l|llll@{}}
\toprule
                      & \multicolumn{4}{c}{Average Sumrate per Network Size}                     \\ \midrule
Algorithm             & 20              & 30               & 40               & 50               \\ \midrule
Max Power             & 62.547          & 79.038           & 96.168           & 106.364          \\
WMMSE                 & 84.563         & 109.972          & 132.677          & 147.463          \\
\midrule
PCGNN                 &84.237           & 109.892          & 131.403          & 147.165          \\
UWMMSE                & 83.081          & 108.335          & 131.048          & 146.035          \\
Graph Transformer   & 85.670          & 111.376          & 133.554          & 148.951          \\
\midrule
Proposed Method         & \textbf{85.968} &\textbf{111.584}  & 134.320  & \textbf{149.592}    \\
Proposed Method: Multi Node & 84.999          & 111.543          & \textbf{134.487}          & 149.343          \\ \bottomrule
\end{tabular}
\label{tab:my-table}
\end{table}

\begin{figure}[tb]
\centerline{\includegraphics[width=0.5\textwidth]{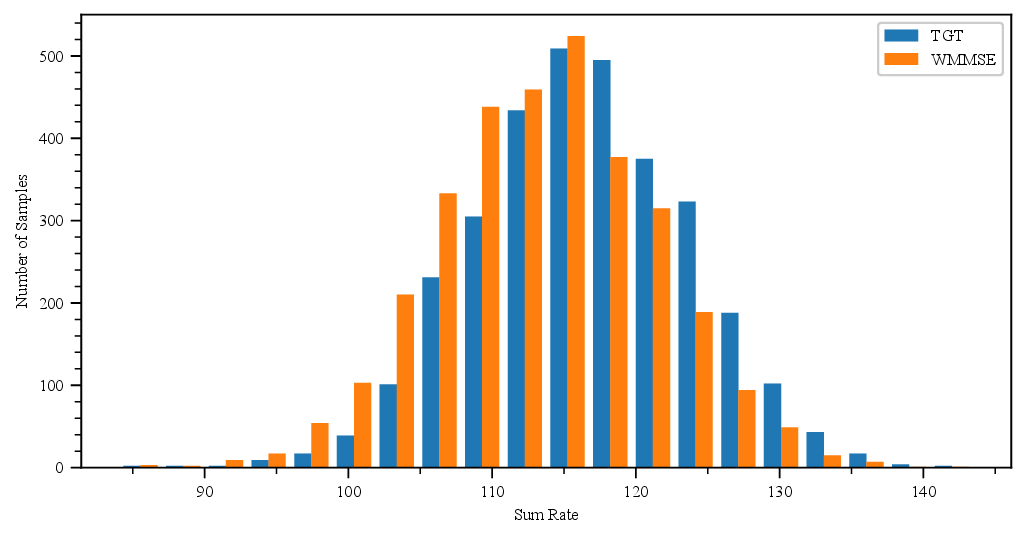}}
\caption{Histogram of sum rates between TGT and WMMSE. Compared across 32000 different channel instances with $n=50, \sigma = 2.6 \times 10^{-5}.$}
\label{fig:per}
\end{figure}
Table \ref{tab:my-table} shows the comparison across various number of pairs. Each data-driven model is trained only on the dataset with the corresponding network size. Multi-node is a single model that was trained on a even distribution of network size of $\{20,30,40,50\}$, with the same number of samples as mentioned above. The proposed method outperforms all methods across various network sizes for both fixed node training and multi-node training. As the training dataset only contains 50 channel instances, we do further comparison of the models on a fixed network topology with 32,000 randomly sampled channel fading. Fig. \ref{fig:per} shows that the proposed method follows a similar distribution curve as WMMSE, with a consistent advantage on all range of channel instances. 
\subsection{Deviation from the Original Network Configuration}
While deep learning models outperform classical algorithms with lower computational complexity, it's performance is only guaranteed within the training data distribution, and thus must be tested with its generality across unseen data distributions to understand when to retrain the model. We test the model's generality across network size, fading distribution, and pair density. We then compare the computational complexity, and provide analysis on the relationship between model size and generality. \par
\begin{figure}[t]
\centerline{\includegraphics[width=0.5\textwidth]{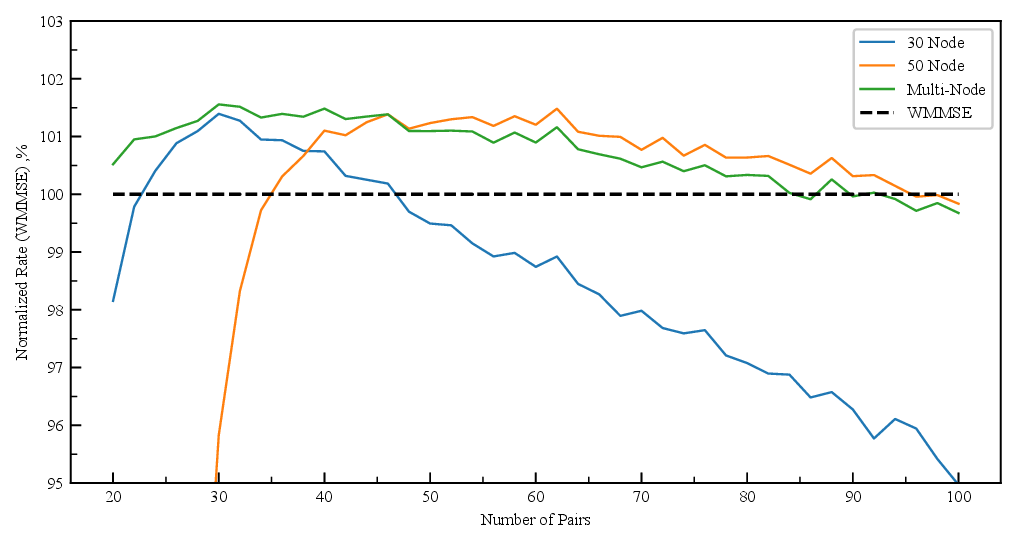}}
\caption{Normalized sum rate of TGT across a wide range of unseen network sizes. $\sigma = 2.6 \times 10^{-5}.$}
\label{fig:gen}
\end{figure}

In Table \ref{tab:my-table}, each model was trained only on a fixed number of $n$. We compare how our model performs across a sweep of network sizes, ranging from 20 to 100 pairs. 10 channel instances are sampled for each 100 random network topologies. Fig. \ref{fig:gen} shows the result across three different plots $n=30,50$ and multi-node model, where the sum-rate is normalized with WMMSE for ease of comparison. A taper of performance is clearly visible from the model's trained network size, but the model is able to outperform WMMSE even outside of the trained pair size. However, we can see that the multi-node model has a more stable performance drop off, performing better on average than fixed size training. This shows the model should be trained on the expected data distribution of the deployed scenario, to prevent frequent retraining of the model.\par

We now consider a scenario where the fading coefficient of the Rayleigh fading changes. Using the same configuration as in Table \ref{tab:my-table},  Table \ref{tab:fade} shows that the proposed model generalizes well from its original fading coefficient 1, outperforming WMMSE up to a fading coefficient of 4 for both fixed model and the multi-node model.  \\
\begin{table}[tb]
\centering
\caption{Model generalization across fading coefficients}
\label{tab:fade}
\resizebox{\columnwidth}{!}{%
\begin{tabular}{@{}lclllllll@{}}
\toprule
\multicolumn{1}{l|}{}          & 0.5   & 1     & 1.5   & 2     & 2.5   & 3     & 3.5   & 4     \\ \midrule
\multicolumn{1}{l|}{Max Power} & 94.9  & 103.4 & 110.4 & 111.8 & 113.9 & 117.5 & 115.2 & 114.5 \\
\multicolumn{1}{l|}{WMMSE}     & 122.2 & 144.3 & 158.2 & 164.2 & 169.4 & 174.9 & 174.3 & \textbf{174.0} \\
\multicolumn{1}{l|}{TGT} & \textbf{123.1} & \textbf{146.3} & \textbf{160.3} & \textbf{165.9} & \textbf{170.2} & \textbf{175.6} & \textbf{174.4} & 173.9 \\ \bottomrule
\end{tabular}%
}
\end{table}
While the model is able to generalize across multiple network size and fading coefficients, the proposed method struggles when the overall density of the network changes from the training data. We test the proposed model on a fixed pair network of $n=50$, but change the the distribution of transmitters from $[-n,n]^2$ to $[-n/r, n/r]^2$, where $r \in [0.25,4]$. Table \ref{tab:den} shows the model struggles to generalize on deviation from the trained pair density. Note that the model still out perform WMMSE when trained on the different density data. \\
\begin{table}[t]
\centering
\caption{Model generalization across network field sizes}
\label{tab:den}
\resizebox{\columnwidth}{!}{%
\begin{tabular}{l|cllllll}
\toprule
          & 200   & 150   & 100   & 50             & 25   & 17   & 12   \\ \hline
Max Power & 217.2 & 210.3 & 179.7 & 107.4          & 44.3 & 22.3 & 14.8 \\
WMMSE & \textbf{221.4} & \textbf{219.3} & \textbf{196.2} & 146.8 & \textbf{96.0} & \textbf{70.0} & \textbf{57.0} \\
TGT      & 215.0 & 216.7 & 195.0 & \textbf{148.7} & 88.9 & 53.3 & 37.6 \\ \bottomrule
\end{tabular}%
}
\end{table}
\subsection{Computational Complexity}
\begin{figure}[tb]
\centerline{\includegraphics[width=0.5\textwidth]{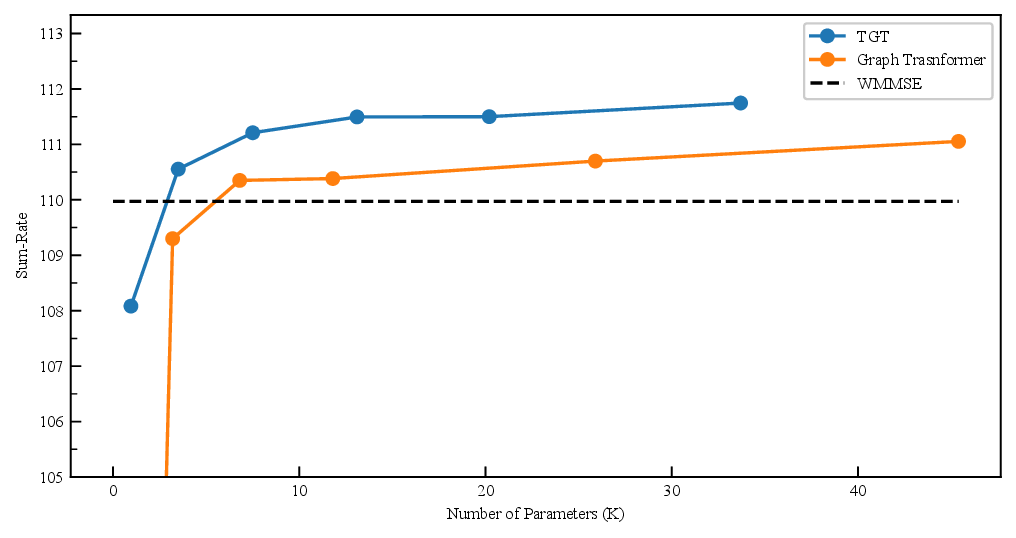}}
\caption{Model performation in relation to parameter size. All models were trained on $n=30$, $\sigma = 2.6 \times 10^{-5}.$ Sum-rate of WMMSE is plotted for reference}
\label{fig:scale}
\end{figure}

\begin{figure}[tb]
\centerline{\includegraphics[width=0.5\textwidth]{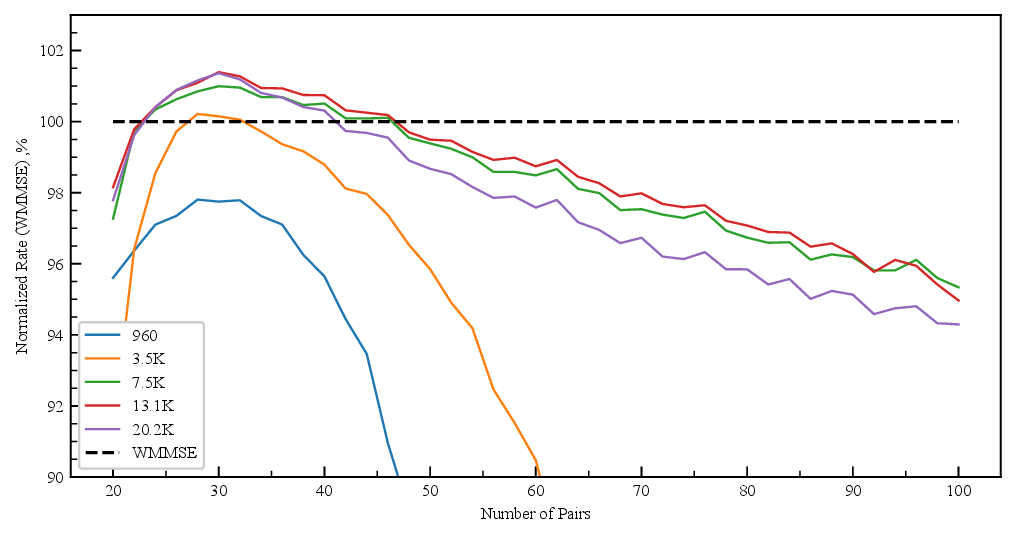}}
\caption{Model generalization in relation to parameter size. Every model was trained on a dataset of $n=30$, $\sigma = 2.6 \times 10^{-5}.$}
\label{fig:scale_gen}
\end{figure}

One advantage of data-driven models over iterative classical algorithms is their simpler computational complexity. WMMSE is known to be run on $O(KN^3)$ time, where N is the number of pairs and K is the number of iterations. Like all other GNN based models, the proposed model has a computational complexity of $O(LFN^2)$, where L is the number of layers, F is the dimension of the linear weights, and N is the number of pairs. Since the computational complexity rely on the model parameter, we compare how the proposed model performs across various number of parameter size. We compare the model with various widths from $d=[4,104]$, in which the model parameter ranges from 960 to 33.7K. Fig. \ref{fig:scale} compares the performance of the proposed model trained on a fixed pair size of $n=30$. For the different parameter sizes, we retain the original ratio of the hidden layers dimensions and scale accordingly while keeping the training conditions same as Section IV-A. The proposed method outperforms WMMSE already at the parameter size of 3.5K, and seems to converge at the parameter size of 13K. Graph transformer achieves equivalent performance with ten times more parameters.\\
While the proposed model outperforms WMMSE at a smaller parameter size, this comes with a price of generality. Fig. \ref{fig:scale_gen} shows that models with smaller parameters fail to generalize to unseen network configurations while models with larger parameters have a slower drop off. Thus, there is a trade-off of computational simplicity and model generality, which decides the frequency of retraining required once the deployed network deviates from its trained data distribution. \\

\section{Conclusion }
We proposed a graph transformer based architecture called TGT for the wireless power allocation problem. It was observed that as the power control problem displays a heterophilous distribution with lower level of noise, architectures such as graph transformer would perform well. With parameter sharing and removing additional non-linear layers, the proposed model is able to achieve state-of-the-art results with a fraction of trainable parameters. Analysis of the model to unseen data distributions shows that although the model is able to generalize on a range of unseen distributions, limits exist to the type of perturbations, which shows a trade-off between model complexity and performance.

\bibliography{reference} 
\bibliographystyle{IEEEtran}
\vspace{12pt}

\end{document}